\documentclass[aps, pra, preprint]{revtex4}
\usepackage[utf8]{inputenc}
\usepackage[T1]{fontenc}
\usepackage[english]{babel}
\usepackage[dvipsnames]{xcolor}
\usepackage{amsmath, mathtools}
\makeatletter
\newcommand{\mathleft}{\@fleqntrue\@mathmargin40pt}
\newcommand{\mathcenter}{\@fleqnfalse}
\makeatother

\usepackage{amssymb}
\usepackage{setspace}
\usepackage{graphicx}
\usepackage[export]{adjustbox}
\usepackage{float}
\begin{document}
\title{Frequency-dependent specific heat in quantum supercooled liquids: A mode-coupling study}
\author{ Ankita Das$^{*}$, Eran Rabani$^{\dagger}$, Kunimasa Miyazaki$^{\ddagger}$, and Upendra Harbola$^{*}$}
\affiliation{ $^{*}$Inorganic and Physical Chemistry, Indian Institute of Science, Bangalore 560012, India.}
\affiliation{$^{\dagger}$ Department of Chemistry, University of California, and Materials Sciences Division, Lawrence Berkeley National Laboratory, Berkeley, California 94720, United States;\\ The Sackler Center for Computational Molecular and Materials Science, Tel Aviv University, Tel Aviv 69978, Israel.  }
\affiliation{$^{\ddagger}$Department of Physics, Nagoya University, Nagoya 464-8602, Japan}
\affiliation{$^{a)}${\bf Author to whom correspondence should be addressed: ankitadas@iisc.ac.in}}
%\date{}

\begin{abstract}
  Frequency-dependence of specific heat in supercooled hard sphere liquid is computed using quantum mode-coupling theory (QMCT).
  Mode-coupling equations are solved using recently proposed perturbative method that allows to study relaxation in the 
  moderate quantum regime where quantum effects assist liquid to glass transition.  
  Zwanzig's formulation is used to compute the frequency-dependent specific heat in supercooled state using dynamical 
  information from QMCT.  Specific heat shows strong variation as the quantumness of the liquid is changed, which becomes 
  more significant as density is increased. It is found that, near the transition point, different dynamical modes contribute to the specific heat in the classical and the quantum liquids.
   
\end{abstract}

\maketitle

\section{Introduction}
 The glass transition has historically been viewed as a second-order phase transition \cite{jp990874b,debendetti} involving a change in the specific heat of supercooled liquids, while the volume and entropy do not show any marked difference during the transition. Liquid-like modes that contribute to the measured thermodynamic quantities  \cite{angell_76} (such as specific heat, thermal expansion coefficient, compressibility etc.) 
near the glass transition are no longer able to equilibrate on the experimental time scales. In order to understand the complex relaxation 
behavior in supercooled liquids, several studies have investigated the response of a supercooled system to  energy fluctuation, known as 
frequency dependence of specific heat in supercooled liquids \cite{PhysRevB.34.1631,particle,PhysRevB.78.144203}.

  The study of specific heat $C(T)$ in supercooled liquids is interesting as it is directly related to Kauzmann paradox \cite{Kauzman1948}. 
  Entropy of the liquid, defined  thermodynamically as the integral $\int dT C(T)/T$, is greater than the crystal entropy at the melting point. As Kauzmann experimentally noted \cite{Kauzman1948}, configurational entropy in supercooled liquids falls rapidly as temperature is decreased and, if continued,  it
  will become less than that of the crystal entropy at finite temperature (so called Kauzmann temperature, $T_K$) which is unphysical. One resolution of the Kauzmann paradox is to say that there must be a phase transition before the configurational entropy of the 
   supercooled liquid becomes lesser than that of crystal. This implies that at a temperature $T(>T_K)$, the dynamics must get arrested, defined as the 
   glass transition temperature ($T_G$). 
    Recent simulation study \cite{Berthier11356} has been able to achieve deeply supercooled states, even lower than what has been realized in  experiments, and successfully confirming the steep decrease in entropy near the glass transition. 

In 1985  Birge and Nagel  \cite{Birge1985} developed an experimental technique to study the dynamic response in supercooled liquids (propylene and glycerol) and obtained specific heat as function of frequency, famously known as ``spectroscopy of specific heat''. In order to understand the origin of the frequency 
 dependence of the specific heat, Oxtoby \cite{1986JChPh..85.1549O} analyzed the Birge and Nagel experiment based on generalized 
 hydrodynamic theory by including a new internal mode relaxing on a time scale of hydrodynamic modes. 
 Later, Zwanzig \cite{1988JChPh..88.5831Z} argued that even without introducing any internal mode the experimental 
 observation of Birge and Nagel could be well explained with generalized hydrodynamics. Zwanzig showed that experimentally what is 
 measured as the frequency-dependent specific heat is related to frequency-dependent longitudinal viscosity. Thus diverging viscosity in supercooled state is related to the frequency-dependent specific heat.
 
 Mode-coupling theory (MCT) is based on generalized hydrodynamic approach which has been used successfully to study relaxation dynamics 
 of glass forming materials using only static property  such as temperature, density, and microscopic structure as 
 input \cite{das_2011_1, Kim-Mazenko,MCT2005, mct_1992}. MCT has been applied to study the  
 relation between dynamics of density fluctuation and frequency-dependent specific heat  in classical supercooled liquids \cite{Harbola2001}.
Although, most of MCT applications have been in the classical regime, there are interesting examples, where 
clear deviations from classical predictions have been attributed to quantum tunneling \cite{zeller_paul_1971, ref_low_sph, 2019_sph}. 
To understand the role of quantum effects on the dynamics of supercooled liquids, quantum mode-coupling theory (QMCT)
has been formulated \cite{PhysRevB.13.3825,R2002}.  In a recent simulation study \cite{Bagchi} on vitrification of supercooled water, it has been 
shown that quantum effects play an important role in determining the frequency-dependent specific heat of water as well as ice. 
 
In the present work we focus to study how the quantum fluctuations affect the frequency-dependent specific heat in supercooled liquids.
We apply QMCT to compute structural relaxation in the moderate quantum regime. Zwanzig's approach is used to compute the 
frequency-dependent specific heat using QMCT results for dynamics as inputs. In order to solve the dynamical QMCT equations we use a recently proposed \cite{das2020structural} perturbation method which allows us to study the dynamics in the supercooled quantum liquids. The quantum effects are found to become increasingly important as the density is increased. At fixed density,  slower modes start to contribute more to the specific heat as the quantumness increases.

In the next section we  define frequency-dependent specific heat and present some essentials of QMCT which will be used in computing the frequency-dependent specific heat. In Sec. III we discuss our results. We conclude in Sec. IV.
 
 \section{Specific heat and QMCT}
 
 In order to define specific heat, we use standard hydrodynamic equations \cite{2013i} where mass density, momentum density and  energy density
  are the dynamic variables of interest. 
As shown by Zwanzig \cite{1988JChPh..88.5831Z}, using generalized hydrodynamic equations for these variables, the following  generalized Fourier heat equation 
can be obtained \cite{Harbola2001},
\begin{equation}
 i\omega\delta T(q,\omega)=-q^2\chi(q,\omega)\delta T(q,\omega),
 \label{eq1}
\end{equation}
where $\delta T(q,\omega)$ is the  temperature fluctuation at wavelength $q$ with frequency $\omega$,
$\chi(q,\omega)=\frac{\kappa}{[\rho_0 c_p(q,\omega)]}$ is thermal diffusivity, $\kappa$ is thermal conductivity, $\rho_0$ is average mass density of the system.  $c_p(q,\omega)$ is the frequency ($\omega$) and wave-vector ($q$) 
dependent specific heat, defined as :
\begin{equation}
 c_p(q,\omega)=c_v\Big[1+(\gamma_q-1)\frac{1}{1+i\omega\Gamma(q,\omega)}\Big],
\label{specific}
 \end{equation}
where $\gamma_q=\frac{c_p(q)}{c_v}$, $c_v$ is specific heat at constant volume and $\Gamma(q,\omega)$ is normalized longitudinal viscosity.
It is to be emphasized that the form of the hydrodynamic Eqs. (\ref{eq1}) and (\ref{specific}) remain same for 
classical and quantum systems \cite{Barth_2018}, the quantum signatures are included through the transport coefficients.
 Equation (\ref{specific}) can be rearranged to define 
normalized frequency-dependent specific heat $C_p(q,\omega)=\frac{c_p(q,\omega)-c_v}{c_p(q)-c_v}=\frac{1}{1+i\omega\Gamma(q,\omega)}$,
so that $C_p(q,0)=1$ for all $q$. 
 For realistic systems such as liquid polyvalent metals like Ga, Cd, Zn,  which can be approximated as hard-sphere (HS) system, $c_v$ lies close to $3NK_B$, with 
 the ratio $c_p/c_v$ in the range ($1.08-1.25$) \cite{cv_ref,ref22}.
Clearly, Eq. (\ref{specific}) suggests that $C_p(q,\omega)$  is complex in nature, where the real part corresponds to the propagation of heat fluctuations through the system and the imaginary part contains information of the important modes that contribute to heat absorption. In order to compute the $C_p(q,\omega)$ we need to evaluate $\Gamma(q,\omega)$ for the quantum system. Viscosity gives a direct measure of the system's resistance to flow when shear stress is applied. Viscosity is intimately related to the relaxation time of density correlation. Slower relaxation of the density correlation is reflected through higher viscosity in the system. The $\Gamma(q,\omega)$ is related to the memory kernel, $M(q,t)$, and  the characteristic frequency of the system, $\Omega_q$ (to be discussed later)  as \cite{Harbola2001}  
\begin{equation}
 \Gamma(q,\omega)=\frac{1}{\Omega_q^2}\int_0^{\infty} dte^{i\omega t}M(q,t).
 \label{10}
\end{equation}
 The real part of $\Gamma(q,\omega)$ provides information of the liquid-like viscous behavior in the system and the imaginary part corresponds to the solid-like elastic property of the system \cite{2013i}. The memory kernel, $M(q,t)$  \cite{das2020structural,Markland2012} is evaluated using QMCT as described below.

 QMCT is used to study the relaxation dynamics in simple liquids  characterized in terms of the relaxation of density correlation function \cite{Markland2012,R2002}
\begin{equation}
C_{\rho\rho}(q,t)=\frac{1}{N}\langle\hat\rho(-q,0)\hat\rho(q,t)\rangle
\end{equation}
where density operator is $\hat\rho(q)=\sum_{i=1}^{N}e^{iq\cdot \hat r_i}$, $N$ is the 
total number of particles, $\hat r_i$ is position vector of the $i^{th}$ particle and $\langle ... \rangle$ denotes 
quantum mechanical ensemble average. 
Due to difficulty in computing directly the full quantum mechanical correlation, ring polymer approach \cite{Chandler1981}
is used to calculate Kubo-transformed correlation function, 
\begin{equation}
 \tilde C_{\rho\rho}(q,t)=\frac{1}{N\beta \hbar}\int_0^{\beta \hbar}d\lambda\langle C_{\rho\rho}(q,t+i\lambda)\rangle,
\end{equation}
where $\beta=\frac{1}{k_BT}$ , $k_B$ is Boltzmann  constant and $\hbar$ is the Planck's constant. 
The Kubo-transformed correlation is related to the quantum correlation via the well known relation \cite{PhysRevE.87.062133}.

Time evolution of $\tilde{C}_{\rho\rho}$ is given in terms of nonlinear integro-differential 
equation given in Refs. \cite{R2002, Markland2012},
\begin{equation}
\frac{d^2{\tilde{ C}}_{\rho \rho}(q,t)}{dt^2}+\Omega_q^2 \tilde C_{\rho \rho}(q,t) +\int_0^t dt^{\prime} M(q,t-t^{\prime})
\frac{d {\tilde{ C}}_{\rho \rho}(q,t^{\prime})}{dt^{\prime}}=0,
\label{1}
\end{equation}
  where $\Omega_q^2=\frac{q^2}{m\beta\tilde S(q)}$, $m$ denotes the particle mass and $\tilde S(q)$ is the zero-time Kubo-transformed density correlation.  
  
In order to quantify quantumness in the system, we use parameter
 $\Lambda^*=\frac{\hbar}{\sqrt{m\sigma^2K_BT}}$ which is the ratio of particle thermal wavelength and its size $\sigma$. 
 The quantum  fluctuations become important  with gradual increase in $\Lambda^*$. At $\Lambda^* \to 0$ limit the QMCT reduces 
 to the classical one \cite{gotze_84}. For short times, dynamics of $\tilde C_{\rho\rho}(q,t)$ is mainly governed by the characteristic frequency $\Omega_q$ 
 of the system. However, as time increases the memory term contributes through renormalizing the viscosity of the system in a non-linear manner; consequently, 
 the dynamics of  $\tilde C_{\rho\rho}(q,t)$ slows down.  

To calculate $\Gamma(q,\omega)$ using Eq. (\ref{10}), we require to solve Eq. (\ref{1})  self-consistently along with the memory function. 
 The memory function has been derived in Ref. \cite{Markland2012} in the frequency domain. 
 However, a direct implementation of the full memory function in computing the supercooled dynamics using self-consistent method
 becomes numerically very challenging. 
To overcome this difficulty, we have recently proposed a perturbative approach \cite{das2020structural} which expresses the memory function 
perturbatively in $\Lambda^*$ (Eq. (11) in ref. \cite{das2020structural}) directly in the time-domain which has been much easier to use in numerical calculations
for moderate quantum regime ($\Lambda^* < 0.1$).

\section{Results}
 %%%%%%%%%%%%%%%%%%%%%%%%%%%%%%%%%%%%%%%%%%%%%%
\begin{figure}[H]
 \begin{center}
 \begin{center}
 \includegraphics[width=12 cm,height=9 cm]{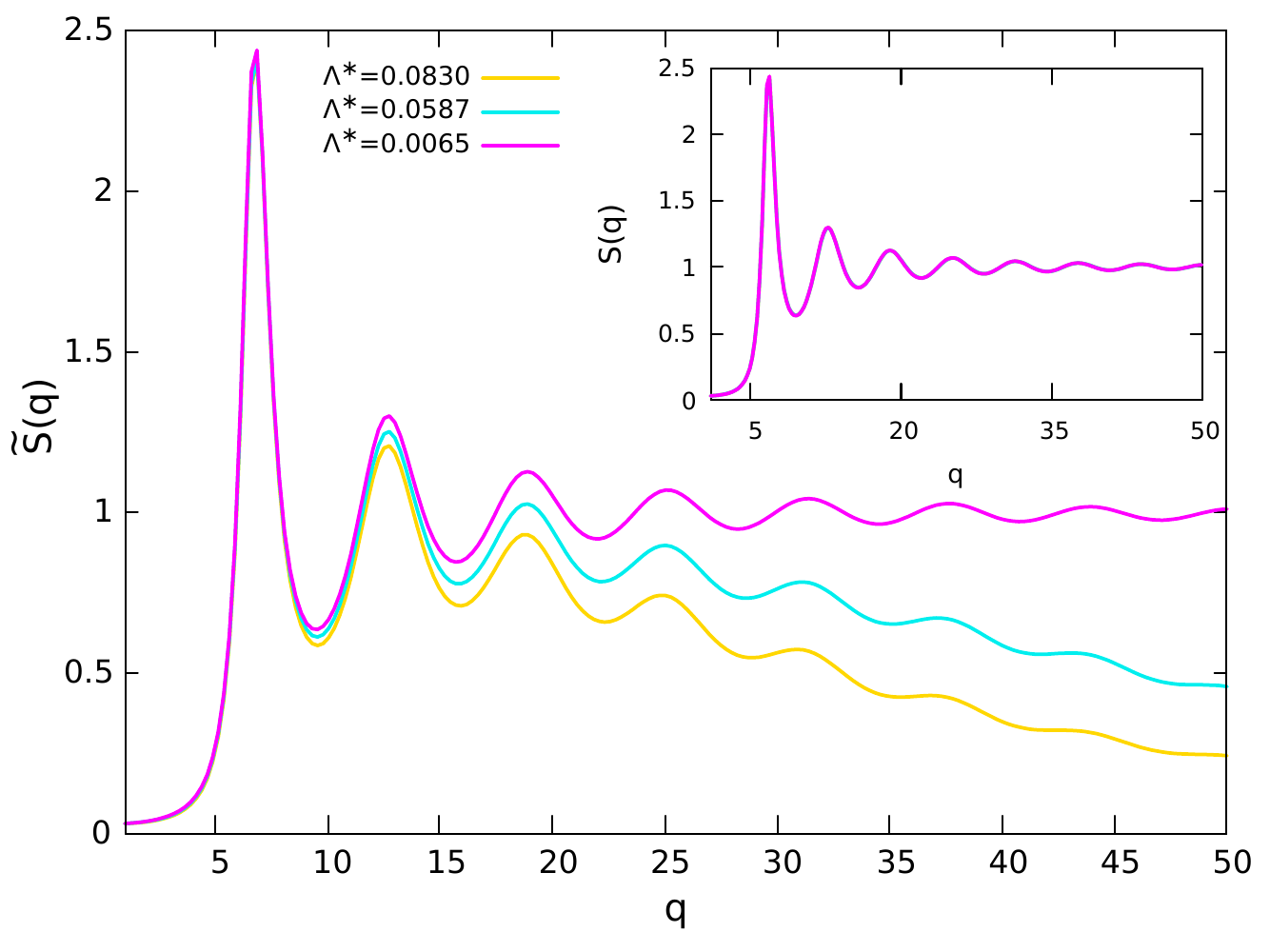}
 % kubo_quantum_structure_factor.pdf: 0x0 px, 300dpi, 0.00x0.00 cm, bb=
\end{center}

 \end{center}
\caption{$\tilde S(q)$ for $\eta=0.445$ for $\Lambda^*=0.0065$, $0.0587$ and $0.0830$. Inset: $S(q)$ for the same $\Lambda^*$ values with 
no significant differences . $q$ is given in the units of $\sigma^{-1}$.}
\label{fig1}
\end{figure}
%%%%%%%%%%%%%%%%%%%%%%%%%%%%%%%%%%%%%%%%%%%%%%%%%%%%%%%%%%%

The QMCT equations are solved self-consistently for a one-component HS system. The static structure factor $S(q)$ for HS supercooled liquid  are generated by solving RISM \cite{2013i} equations with PY \cite{physrev.110.1} closure. $\tilde S(q)$, which is needed to solve 
Eq. (\ref{1}), is generated using the approximation \cite{Markland2012}, $\tilde S(q)\approx \frac{2S(q)}{\beta\hbar\Omega_q\Delta n(\Omega_q)} $, where $\Delta n(\Omega_q)=n(\Omega_q)-n(-\Omega_q)$ and the Bose distribution function $n(\Omega_q)=\frac{1}{e^{\beta \hbar \Omega_q}-1}$, shown in  Fig. (\ref{fig1}) for different $\Lambda^*$ at volume-fraction $\eta=0.445$. 
Note that for the classical HS system ($\Lambda^*=0$) the transition point, $\eta_c=0.509$. 
As $\Lambda^*$ increases the $\eta_c$ decreases \cite{Markland2011} as for $\Lambda^*=0.1016$,  $\eta_c=0.445$  \cite{das2020structural}. In the inset we show $S(q)$  for  the same $\Lambda^*$ values for comparison with $\tilde S(q)$.  
 Compared to the quantum structure factor, $\tilde S(q)$  varies significantly with $\Lambda^*$; 
 $\tilde S(q)$  decreases at larger $q$-values due to increasing quantum uncertainty in the particle position as $\Lambda^*$ is increased, 
 while $S(q)$ approaches to $1$.

%%%%%%%%%%%%%%%%%%%%%%%%%%%%%%%%%%%%%%%%%%%%%%%%%%%%
\begin{figure}
\begin{center}
 \includegraphics[width=10 cm,height=8 cm]{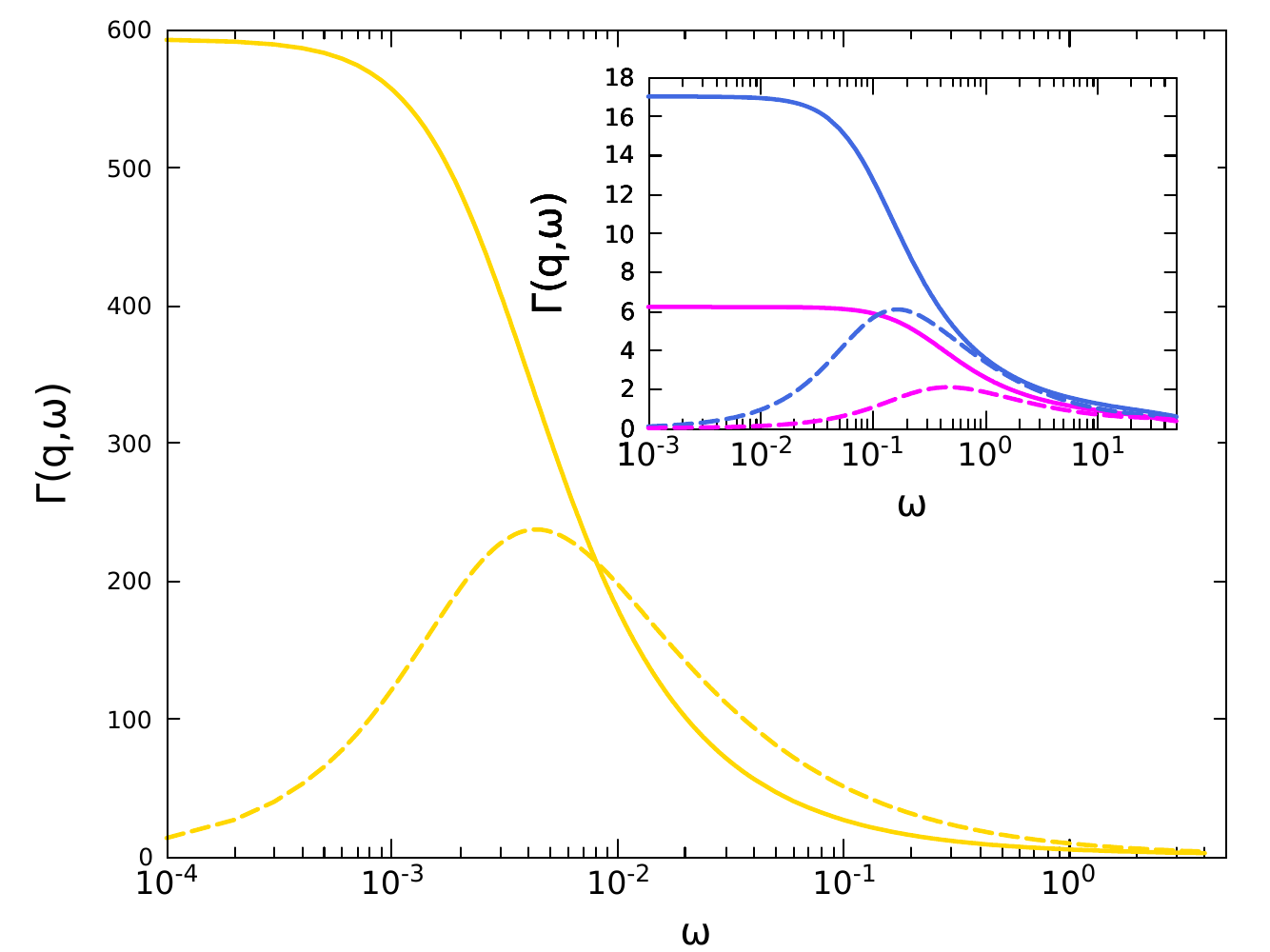}

\end{center}

\caption{Real (solid curve) and imaginary (dashed curve) parts of $\Gamma(q,\omega)$ are shown for $\Lambda^*=0.095$ (yellow) at $q=7.11$ with frequency (in units of $1/\tau$). Inset shows the real and imaginary part of  $\Gamma(q,\omega)$ for $\Lambda^*=0.0587$ (blue) and $\Lambda^*=0.0065$ (pink) at $q=7.11$.}
\label{fig2}
 \end{figure}

%%%%%%%%%%%%%%%%%%%%%%%%%%%%%%%%%%%%%%%%%%%%%%%%%

$S(q)$ and $\tilde S(q)$ are used as inputs to solve the dynamic Eq. (\ref{1}) to compute the normalized longitudinal viscosity, $\Gamma(q,\omega)$ 
defined in Eq. (\ref{10}).
In Fig. (\ref{fig2}), the real (solid curve) and imaginary (dashed curve) parts of $\Gamma(q,\omega)$ are plotted  against frequency $\omega$ (in units of $1/\tau$, where $\tau=\sqrt{m\sigma^2\beta}$) at $q=7.11$ 
(first  peak in $S(q)$) for $\eta=0.445$ and $\Lambda^*=0.095$. For comparison, results for   $\Lambda^*=0.0587$ (blue) and $\Lambda^*=0.0065$ (pink) are shown in the inset. An increase in $\Lambda^*$ leads to increase in viscosity  which results in slower relaxation of density fluctuations \cite{das2020structural}. The peak position  in Im$\Gamma(q,\omega)$ shifts to low frequency regime indicating that the system is able to sustain shear stress at low frequencies and respond in a solid-like manner. The value of Re$\Gamma(q,\omega=0)$ rises with increasing $\Lambda^*$ showing an increase in viscosity and a tendency to approach the liquid-glass transition.

%%%%%%%%%%%%%%%%%%%%%%%%%%%%%%%%%%%%%%%%%%%%%%%%%%%%%%%%%%%%%%%%%
\begin{figure}[b]
\begin{center}
 \includegraphics[width=12 cm,height=9 cm]{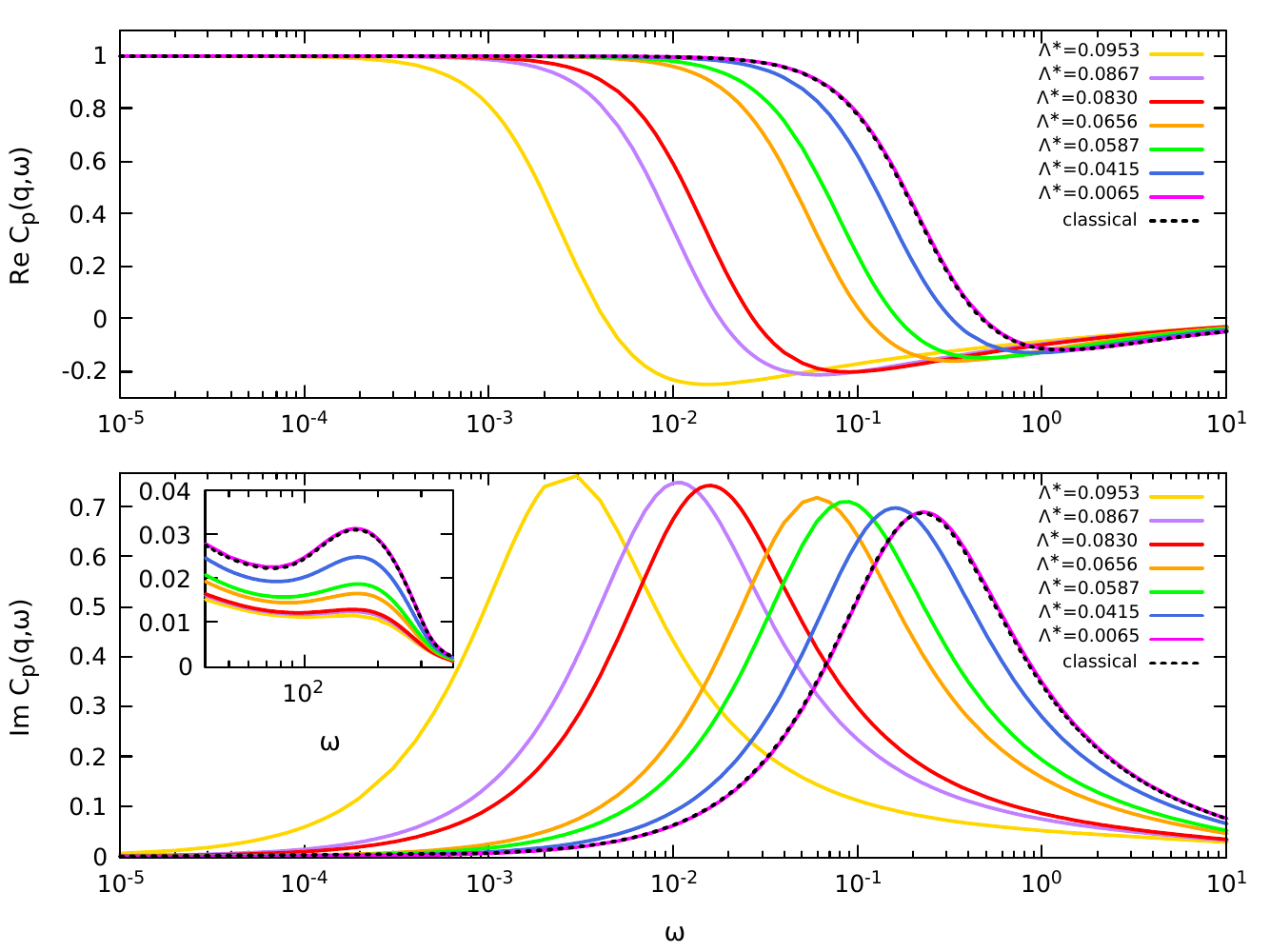}
 
\end{center}

\caption{Real (upper panel) and imaginary parts (lower panel) of frequency-dependent specific heat of quantum HS supercooled 
liquid at volume-fraction $\eta=0.445$ and $q=7.11$ is plotted with frequency 
(in units of $1/\tau$) for $\Lambda^*=0.0953$ to $\Lambda^*=0.0065$ (from left to right). The dotted curve represents specific heat of the classical ($\Lambda^*=0.0$)
HS supercooled liquid which overlaps with the quantum result for $\Lambda^*=0.0065$ (pink). With increasing $\Lambda^*$ the
peak of the imaginary part in specific heat shifts towards the lower-frequency regime. 
Inset in the lower panel shows a secondary peak in Im$C_p(q,\omega)$ around high frequency window ($150-300$). }
\label{fig3}
\end{figure}

%%%%%%%%%%%%%%%%%%%%%%%%%%%%%%%%%%%%%%%%%%%%%%%%%%%%%%%%%%%%%%%%%%%%

We use these viscosity results to compute the specific heat using Eq. (\ref{specific}). 
To analyze the specific heat spectra, the real 
and the imaginary parts of the frequency-dependent specific heat are plotted in Fig. (\ref{fig3}) for different degrees of quantumness. The real and the imaginary 
parts of specific heat, respectively, give information on how an energy fluctuation propagates and relaxes (dissipates) in the liquid. A peak in the imaginary 
part at a certain frequency denotes time-scale of the modes contributing most to the energy dissipation. These are the modes responsible for a non-zero specific heat 
of the supercooled liquid as most of the heat is used in the excitation of those modes. The frequency dependence of specific heat corresponding to $\Lambda^*=0.0065$  is similar to that of the classical one (shown with black dots) and represents the classical limit of quantum specific heat. The peak position in the imaginary part of the specific heat, which corresponds to the relaxation process, 
shifts to the lower frequency values with increasing degree of quantum fluctuation, indicating an increasing 
contribution coming from the slower modes. Thus similar to the classical case when the density is increased, in quantum case too, the higher frequency modes tend to freeze first as the quantumness is increased in the supercooled regime. Inset in the lower panel shows  appearance of secondary peak at high frequency regime $150\leq \omega\leq 300$. The high frequency secondary peak in the specific heat results from the short time dynamics dominated by the characteristic sound mode of frequency $\Omega_q$ \cite{das_2011_1}. As the characteristic frequency $\Omega_q$ at $q=7.11$  does not vary with the quantumness of the system, the secondary peak position remains invariant with changing $\Lambda^*$.

%%%%%%%%%%%%%%%%%%%%%%%%%%%%%%%%%%%%%%%%%%%%%%%
\begin{figure}[H]
\begin{center}
 \includegraphics[width=10 cm,height=8 cm]{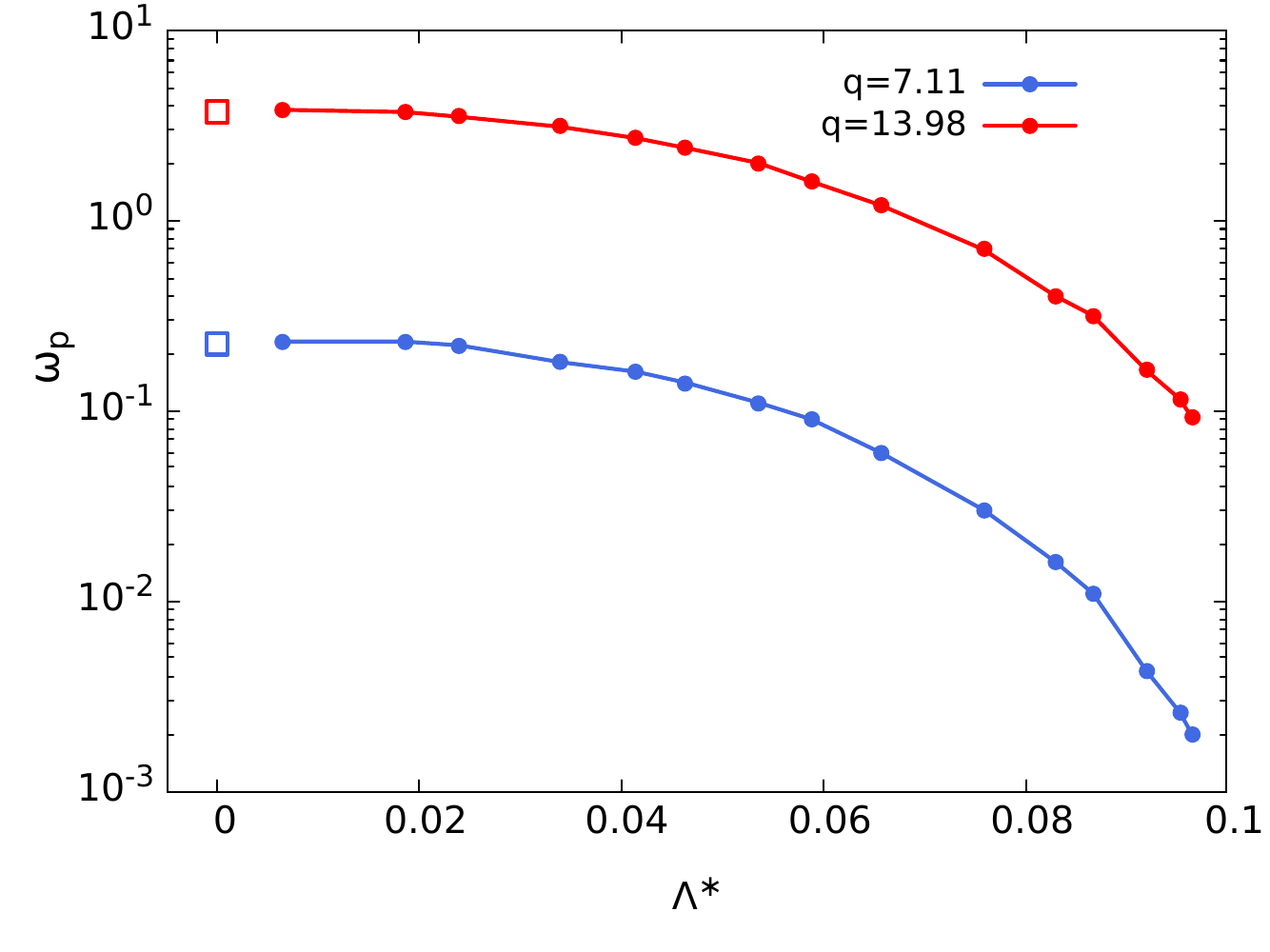}
 % peak_position1.pdf: 0x0 px, 300dpi, 0.00x0.00 cm, bb=
\end{center}

\caption{Peak position ($\omega_p$) in the imaginary part of the frequency-dependent specific heat for supercooled liquid at $\eta=0.445$ for varying 
degree of quantumness ($\Lambda^*$). The red and the blue curves represent shifts in the peak position  for $q=7.11$ and $q=13.98$, respectively. Red and blue void-squares at $\Lambda^*=0$ denote results for the classical liquid.  }
\label{fig5}
\end{figure}
In Fig. (\ref{fig5}) the peak position ($\omega_p$)  of specific heat is plotted with increasing degree of 
quantumness for two different $q-$values. For  
both $q-$values, $\omega_p$ shifts towards lower frequency regime as $\Lambda^*$ increases. Different frequency modes contribute to the specific heat at different length scales. The modes contributing at $q=7.11$ have frequency approximately an order of magnitude lower than modes contributing at $q=13.98$ (second  peak in $S(q)$). 
Similar to the classical case \cite{Harbola2001}, in the quantum case also there is shift in the peak position of the specific heat towards the lower frequencies as the transition point is approached. It is important to compare the nature of this $\omega_p$ shift in quantum and classical supercooled liquids. We compute the shift in $\omega_p$ in both classical and quantum liquids by varying the density.

%%%%%%%%%%%%%%%%%%%%%%%%%%%%%%%%%%%%%%%%%%%%%%%%%%%%
\begin{figure}[H]
 \begin{center}
 \includegraphics[width=11 cm,height=10 cm]{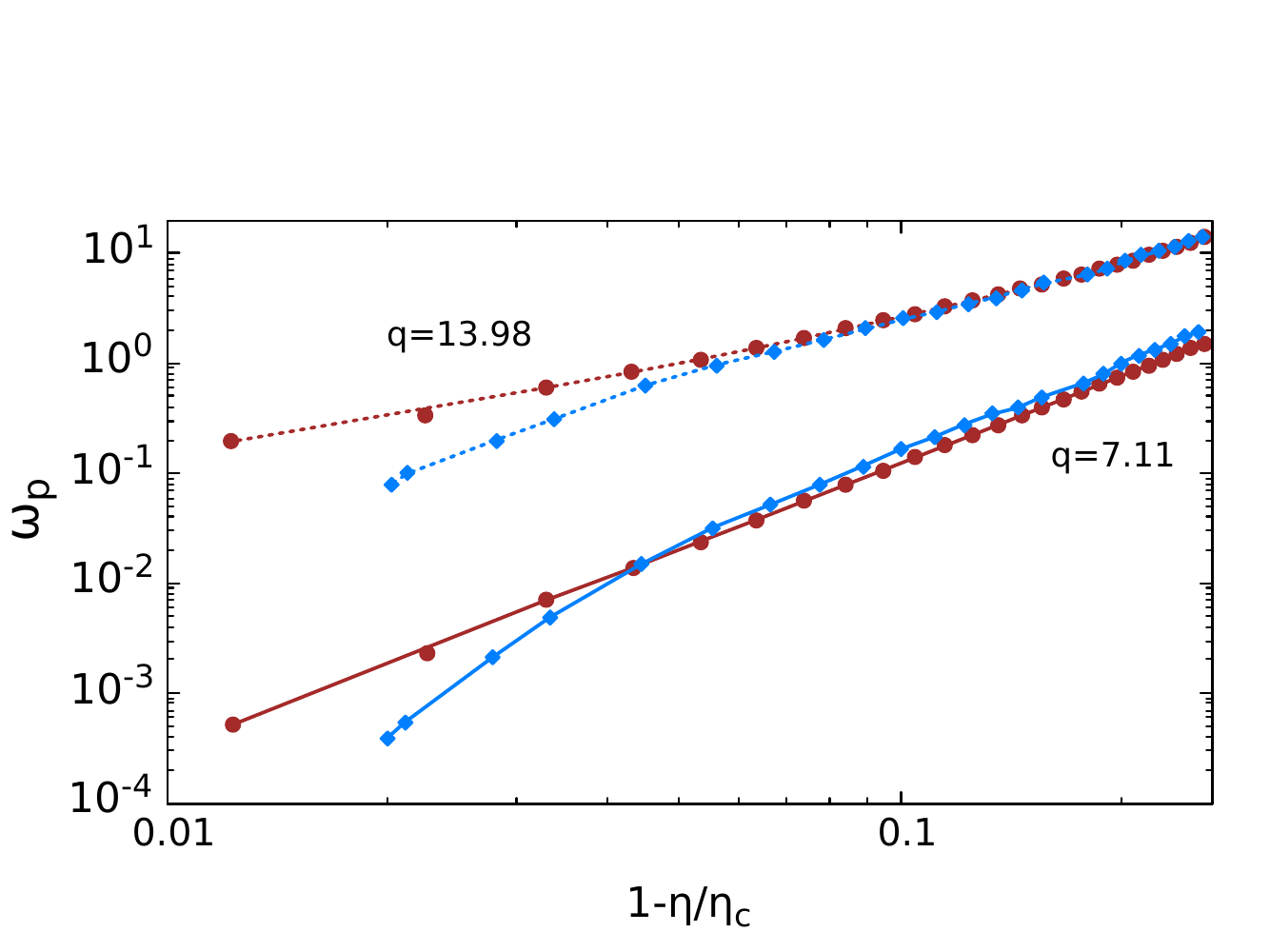}
\end{center}
\caption{The peak position ($\omega_p$) in the imaginary part of the specific heat for supercooled liquid as
 function of  $1-\frac{\eta}{\eta_c}$. Brown and blue curves represent results for the classical and the quantum liquids ($\Lambda^*=0.083$), respectively,
  at $q=7.11$ (bold curve) and at $q=13.98$ (dashed curve).}
\label{fig7}
\end{figure}
%%%%%%%%%%%%%%%%%%%%%%%%%%%%%%%%%%%%%%%%%%%%%%%%%%%%%%%%%%%%%%

 In Fig. (\ref{fig7}), we compare results for the classical (brown) and quantum (blue) liquids for wave-vectors $q=7.11$ (bold curve) and $q=13.98$ (dashed curve). At lower densities (far from the transition point), $\omega_p$ for both the classical and quantum liquids has similar values and shows similar decreasing trend as the density is increased.  However, close to the transition point the decreasing trend in $\omega_p$ for the quantum liquid is qualitatively different from the classical case and shows much faster decay as the density is increased, indicating dominant dynamical quantum effects at higher densities.

%%%%%%%%%%%%%%%%%%%%%%%%%%%%%%%%%%%%%%%%%%%%%%%%%%%%

\section{Conclusion}
Quantum mode-coupling theory is used to study the behavior of frequency-dependent specific heat in supercooled HS quantum liquids. 
The quantumness of the system strongly influences  dynamical relaxation in supercooled liquids. 
The frequency dependence of  the specific heat is found to show significant quantum effects. The slower modes contribute more to 
the frequency dependence as quantumness is increased. This is the trend seen in the moderate quantum liquid and is consistent with the dynamics 
observed for density fluctuations. At low densities, similar frequency modes contribute to the specific heat in both the classical and the quantum supercooled liquids. Quantum signatures become more pronounced as the density is increased. Near the transition point, different frequency modes contribute to the specific heat in the two cases, modes contributing to the specific heat in quantum liquids are much slower compared to those contributing in case of classical liquid.
The detailed study on the quantum effects for simple HS model system presented in this work can serve as a good guide for studying the 
frequency dependence of specific-heat in real systems, for example, low mass colloidal systems (which show hard-sphere like behavior), where quantum 
effects play important role at low temperatures.

In the present calculation, the quantum effects are included using perturbative method valid in the moderate quantum regime. 
It will be interesting to explore the dynamical effects also in the strong quantum regime using non-perturbative approach. 
We plan to study this in the future. 

\section*{Acknowledgments}
AD acknowledges support from University Grants Commission (UGC), India. UH acknowledges Science and Engineering Research Board, India for 
support under the Grant No. CRG/2020/001110, and  TATA Trust Travel Fund from IISc. 
KM acknowledges support by Japan Society for the Promotion of Science (JSPS) KAKENHI
(No.~16H04034   and  20H00128).

\section*{Data Availability }
The data that support the findings of this study are available from the corresponding author upon reasonable request.

\bibliography{ref1}

\end{document}